\documentclass[%
reprint, 
 amsmath,amssymb,
 aps,
]{revtex4-2}

\usepackage{graphicx}
\usepackage{bbm}

\usepackage{dcolumn}
\usepackage{bm}
\usepackage[mathlines]{lineno}
\usepackage{tabstackengine}
\setstackEOL{\cr}
\usepackage{soul,color}

\usepackage{amsmath}

\DeclareMathOperator*{\argmin}{arg\,min}

\begin{document}

\preprint{DRAFT 2}

\title{Machine learning from limited data: Predicting biological dynamics under a time-varying external input}

\author{Hoony Kang}
\author{Keshav Srinivasan}
\author{Wolfgang Losert}

\affiliation{University of Maryland, College Park, Maryland 20742}

\date{\today}

\begin{abstract}


Reservoir computing (RC) is known as a powerful machine learning approach for learning complex dynamics from limited data. Here, we use RC to predict highly stochastic dynamics of cell shapes. We find that RC is able to predict the steady state climate from very limited data. Furthermore, the RC learns the timescale of transients from only four observations. We find that these capabilities of the RC to act as a dynamic twin allows us to also infer important statistics of cell shape dynamics of unobserved conditions.


\end{abstract}

\maketitle

\section{Introduction}\label{introduction}

Machine learning has been applied extensively in analyzing and predicting complex dynamics. It is particularly important for biological systems since they cannot be fully captured with simple mathematical models, and since biological systems are generally non-stationary, i.e. continuously evolve their behavior in time. One Machine Learning architecture known to learn rapidly enough to deal with such time varying data is Reservoir Computing (RC) \cite{jaeger_rc,maass_lsm,chaos,chaos_neuron}. 

In this paper, we implement a RC, specifically an echo state network \cite{jaeger_esn}, to analyze and predict the boundary motion of a large migratory cell that is subject to a time varying electric field $\mathbf{E}$.  We provide this context to the RC by equipping it with a dynamic input channel that supposes a parameter of the dynamical system\cite{kong,xiao}. This allows us to assess the reservoir's predictability for each parameter value, and its adaptability as the parameter is switched mid-prediction. 

We also implement a parallel architecture that incorporates knowledge of the input network structure to improve performance\cite{keshav,parallel_pathak}.


We use the RC with its additional parameter channel as a `dynamics twin' (i.e. as a simulated system that can be used to explore changes in parameters or previously unobserved conditions of the dynamic system) for the $\mathbf{E}$-guided cell boundary motion.  The dynamic twin allows us to (1) estimate the degree to which the measured boundary motion of a cell is deterministic versus stochastic, (2) assess transition characteristics for sudden changes in parameters, and (3) obtain the statistics of interpolated system states.


\section{Model and data background}

The equation describing the node states $\mathbf{r}(t)$ in a reservoir is
\begin{align}
\begin{split}
\mathbf{r}(t+\Delta t)=&\alpha \mathbf{r}(t)+(1-\alpha)\tanh\big[\mathbf{A}\mathbf{r}(t)+\mathbf{W}_{\text{in}}\mathbf{u}(t)\\&+b\mathbbm{1}\big]
\end{split}
\end{align}

where the activation function $\tanh[\cdot]$ is taken to apply to each component of the vector equation in its argument separately.

The strengths of the nonzero elements are set by `hyperparameters' that scale the matrices' elements. Elements of $\mathbf{W}_{\text{in}}\in \mathbb{R}^{\text{N}_\mathbf{r}\times \text{N}_\mathbf{u}}$ are chosen from a uniform distribution, i.e. $\mathbf{W}_{\text{in}_{ij}} \in [-\sigma, \sigma]$ where $\sigma \in \mathbb{R}^+$ is the input scaling. The sparse adjacency matrix $\mathbf{A}\in \mathbb{R}^{\text{N}_\mathbf{r}\times \text{N}_\mathbf{r}}$ is scaled by its spectral radius $\rho$ to ensure the echo state property. $\text{N}_{\mathbf{r}}$ is the number of nodes in the reservoir and $\text{N}_\mathbf{u}$ is the input dimension of $\mathbf{u}$. $b$ is some constant bias, and $\mathbbm{1}$ denotes a vector of 1s.

The state of the reservoir is then cast back onto the original spatial domain of the input data via a linear mapping  $\tilde{\mathbf{u}}:\mathbb{R}^{\text{N}_\mathbf{r}}\rightarrow \mathbb{R}^{\text{N}_\mathbf{u}}$ defined by
\begin{align}
\tilde{\mathbf{u}}(t)=\mathbf{W}_{\text{out}} \mathbf{r}(t)
\end{align}
\noindent where $\mathbf{W}_{\text{out}}\in \mathbb{R}^{\text{N}_\mathbf{u}\times\text{N}_\mathbf{x}}$ is the output matrix that maps the reservoir states back onto the input domain. As $\tilde{\mathbf{u}}(t+\Delta t)$ assumes the prediction of $\mathbf{u}_\text{data}$ after one time-step $\Delta t$, the goal is to obtain the matrix $\mathbf{W}_{\text{out}}$ such that $\tilde{\mathbf{u}}(t)$ best approximates $\mathbf{u}(t)$ for chosen hyperparameters. 

To do this, the reservoir is initiated into a `listening phase' during the first few time-steps of feeding in the input, i.e. for $\tau_\text{listen}$ steps of $\Delta t$. Here, $\mathbf{u}(t)=\mathbf{u}_\text{train}(t)$, thereby entraining the states $\mathbf{r}(t)$ to the dynamics of the input data. As this stage is used solely to synchronize the nodes with the data, these reservoir states are discarded and not used for training.

Upon completion, the reservoir enters the `training phase' for $\tau_\text{train}$ steps of $\Delta t$. The now-synced states are again entrained by $\mathbf{u}(t)=\mathbf{u}_\text{train}(t)$, but all the reservoir states during this phase $\mathbf{r}(t_0+\Delta t)$, $\mathbf{r}(t_0+2 \Delta t)$, ..., $\mathbf{r}(t_0+\tau_\text{train} \Delta t)$ are recorded. For notational brevity, $t_0=\tau_\text{listen}\Delta t$ marks the end of the listening period. We then obtain the trained output matrix $\mathbf{W}_{\text{out}}$ using the recorded reservoir states via ridge regression by minimizing the loss:
\begin{align}
\begin{split}
\argmin_{\mathbf{W}_\text{out}}\bigg[\sum_{t=t_0+\Delta t}^{t_0+\tau_{\text{train}}\Delta t}||&\mathbf{W}_{\text{out}}\mathbf{r}(t)-\mathbf{u}_\text{train}(t)||^2\\
&+\beta \,\text{tr}(\mathbf{W}_\text{out}\mathbf{W}_\text{out}^\text{T})\bigg]
\end{split}
\end{align}

where  $\beta$ is the Tikhonov parameter.

Equipped with the trained $\mathbf{W}_\text{out}$, we are ready to start the prediction stage. To do this, the first few frames of the test data $\mathbf{u}_\text{test}$ is first sent into the reservoir for another listening period in order to re-synchronize the reservoir states $\mathbf{r}(t)$ to the new data. The last frame of the test data in the listening period $\mathbf{u}_\text{test}(t_0)$ is now used as the initial condition to generate the first prediction, $\mathbf{\tilde{u}}(t_0+\Delta t)$. This output is then fed back as the new input to obtain the reservoir's next state to form a \textit{closed} prediction for the next time-step. This process is iterated until the reservoir has advanced up to our desired prediction window.\\

\textit{Parameter tagging}. Per \cite{kong}, we insert an external parameter channel by adding an additional term into the activation function:
\begin{align}
\begin{split}
\mathbf{r}(t+\Delta t)=&\,\alpha \,\mathbf{r}(t)+(1-\alpha)\tanh \big[\mathbf{A}\mathbf{r}(t)+\mathbf{W}_{\text{in}}\mathbf{u}(t)\\
&+\mathbf{W}_{\mathbf{p}}\,\mathbf{p}(t+\Delta t)+b\mathbbm{1}\,\big]
\end{split}
\end{align}

Here, $\mathbf{p}$ supposes some parameter of the system, e.g. an internal bifurcation parameter or of some external forcing. $\mathbf{p}$ is not fed back into the reservoir, as its value is preset. With the temporal dependence of $\mathbf{p}(t)$ , we can teach the reservoir to predict the dynamics corresponding to each value of $\mathbf{p}$. In a dynamical systems theory framework, this external bias parameter assists the reservoir in partitioning its phase space such that dynamics corresponding to each $\mathbf{p}$ reside in distinct regions. Furthermore, we can observe transitions in dynamics when this parameter value switches in real-time, which corresponds to how the reservoir crosses the boundaries between regions of different dynamics in its phase space.

In this paper, we consider a system where $\mathbf{p}(t)$ is homogeneous across inputs ($\mathbf{p}(t)=p(t)\mathbbm{1}$). Note that we use the parameter value at $t+\Delta t$ in our activation function, rather than at $t$, as it should reflect the updated reservoir state $\mathbf{r}$ at $t+\Delta t$.


Following $\mathbf{W}_\text{{in}}$'s construction, the elements of the external parameter input matrix $\mathbf{W}_{\mathbf{p}} \in \mathbb{R}^{\text{N}_\mathbf{r}\times\text{N}_\mathbf{x}}$ are chosen from a uniform distribution such that $\mathbf{W}_{\mathbf{p}_{ij}} \in [-\sigma_\mathbf{p}, \sigma_\mathbf{p}] \,\forall \,(i, j)$ where the external parameter input scaling $\sigma_\mathbf{p} \in  \mathbb{R}^+$ is another hyperparameter.\\

\begin{figure}
    \centering
    \includegraphics[scale=.18]{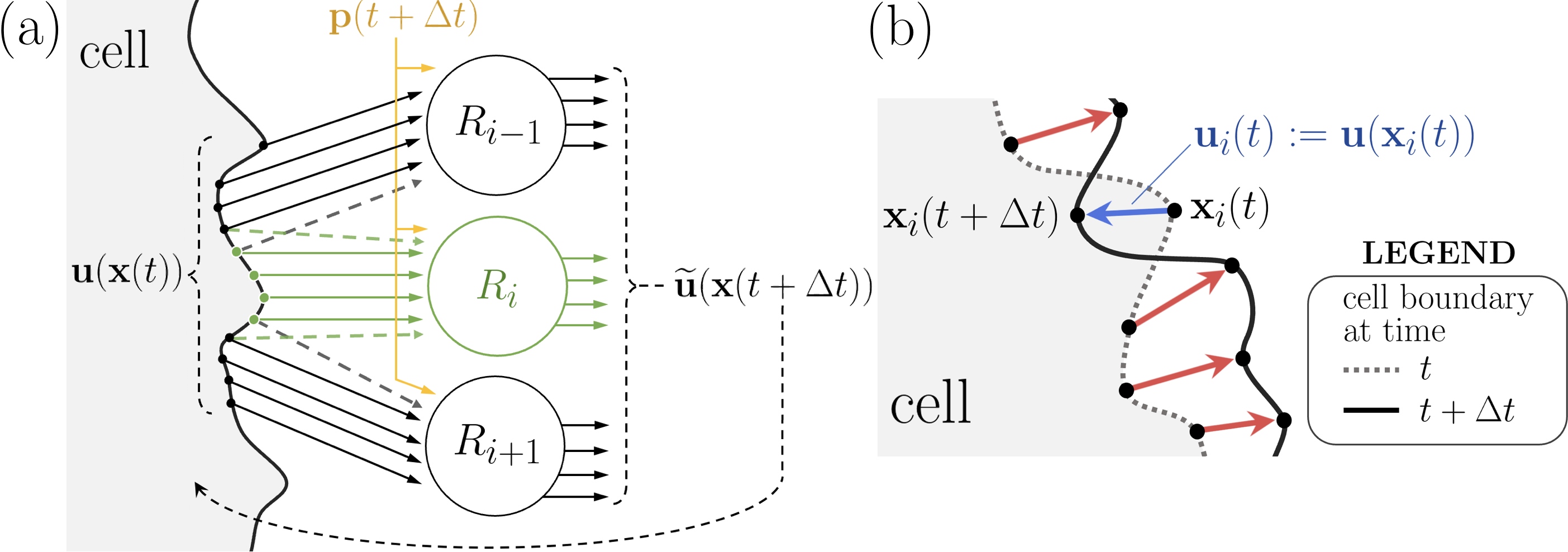}
    \caption{(a) Schematic of the parallel ML architecture with the external parameter channel. Each reservoir predicts a subset of inputs while taking in neighboring inputs (here, set to 1 for visual simplicity) during both training and prediction. All reservoirs take in the parameter channel. (b) Schematic of boundary motion data processing. Each boundary point $\mathbf{x}$ was tracked in time to obtain their respective speeds $\mathbf{u(x)}(t))$ via optical flow analysis. If the point recedes into (protrudes out of) the cell body, its speed is assigned a negative (positive) value, as shown by the blue (red) arrows.}
    \label{fig:schematic}
\end{figure}

\textit{Parallel RC.} As the input system size increases, the reservoir encounters difficulty predicting all the time series simultaneously. It must not only find the proper network structure between all the inputs, but also will waste resources in forming connections between inputs that we know \textit{a priori} should not be coupled. To mitigate this, we can incorporate our knowledge of the input network structure by implementing a series of smaller reservoirs that each take in and predict only a portion of the inputs in parallel, whilst using the closest neighboring inputs to influence each reservoir's evolution (fig. \ref{fig:schematic}a). We call this the `parallel' architecture henceforth. By prohibiting the reservoirs from coupling inputs that are far removed in input network space, the reservoirs seeks to more closely mimic the input coupling structure. This architecture has demonstrated large improvements in predictability\cite{keshav} if the coupling structure is known  \textit{a priori} or can be accurately inferred \textit{a posteriori}.

\begin{figure}
    \centering
    \includegraphics[scale=.13]{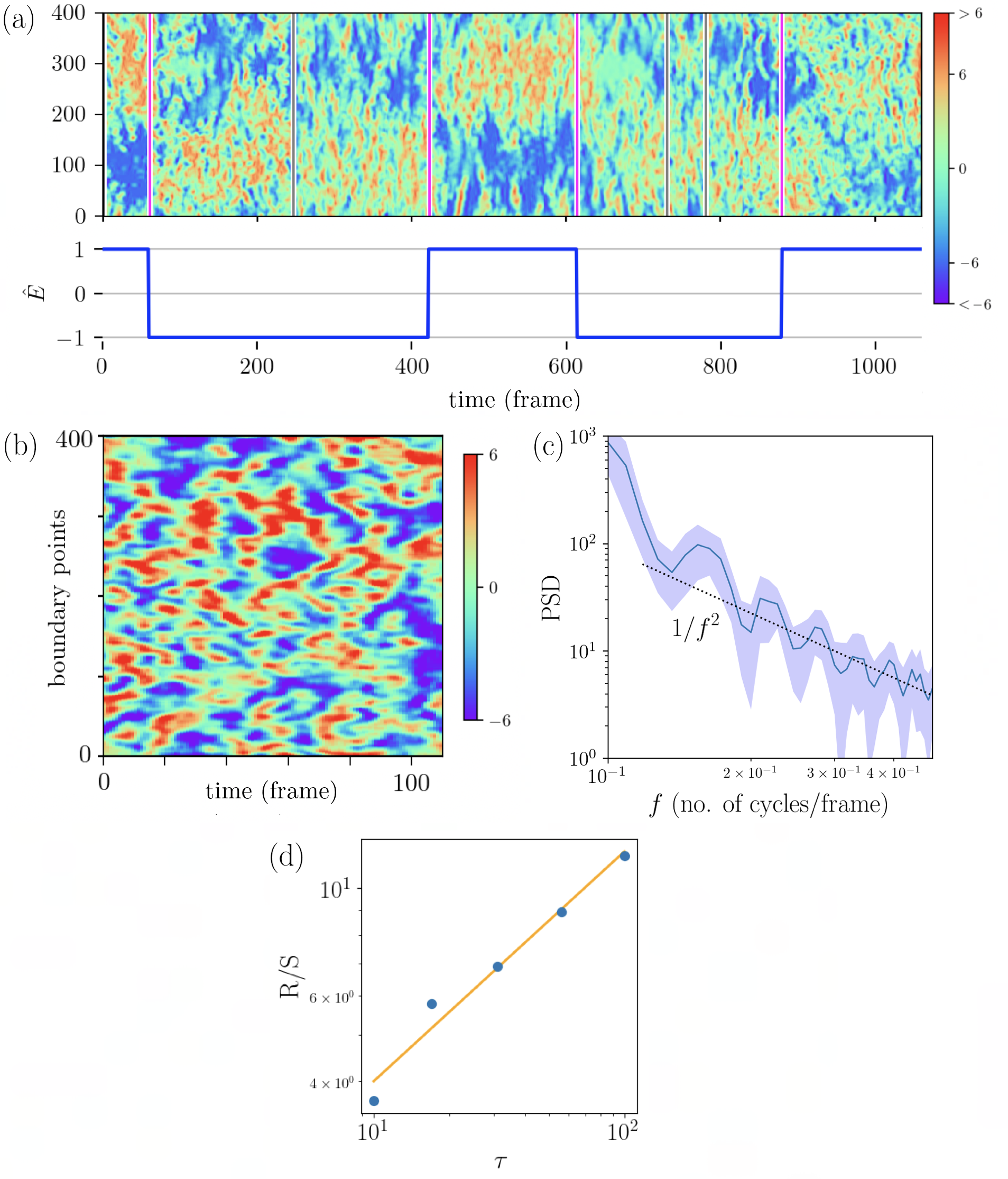}
    \caption{(a) Smoothed boundary motion data used for listening and training. (b) $\hat{E}=0$ data. (c) Power spectral density (PSD) of the data in (b), plotted in logarithmic scales. The solid line is the mean PSD across all 400 boundary points, with 2 standard deviations shaded in to show the PSD distribution across the points. The dashed line shows the theoretical $1/f^2$ power law of Brownian noise. The corresponding $\hat{E}$ direction is shown in blue, and demarcated in magenta lines in the kymograph. Gray lines demarcate splicing points of either different experiments or due to removing experimental errors. (d) A rescaled range analysis plot showing the rescaled range series (R/S) against the time window $\tau$ as defined in eq. \ref{hurst} in logarithmic scales, for the steady state $t\in[300,400]$. The Hurst exponent is given by the slope of the best fit line (for this particular trace, $H=0.47$).}
    \label{fig:data}
\end{figure}
\subsection{Data processing}
We assess the model’s predictability on boundary motion data of a giant Dictyostelium cell lying on a ridged surface that is subject to electrotaxis, generated in ref. \cite{qixin}. The cell was subjected to a spatially homogeneous electric field in a prescribed direction, causing collective migration of the cell along the electric potential gradient with finger-like protrusions of the boundary characteristic of actin-driven motion \cite{qixin,nir,giannone}. Then, the direction of the applied field was flipped after some arbitrary duration, causing the cell to switch directions in movement. The cell’s boundary was discretized to 400 points, with their motions tracked per frame (10 fps) to generate the data sets. The instantaneous velocities of every point were calculated via optical flow analysis, as detailed in \cite{lee}. A schematic is illustrated in fig. \ref{fig:schematic}b.  

Three of four experimental datasets were conjoined to form one longer training data, with a listening stage invoked at every concatenation time point fig.\ref{fig:data}a. The fourth experiment was used as test data to evaluate the performance of the trained RC. 
The data suffers from intensity clipping past ±6 and exhibits a high degree of noise due to experimental design. In order to mitigate these artifacts that may be detrimental to training, we apply a mild Savitzky-Golay filter to interpolate dynamics within these plateaus of saturated intensity and to smooth the noise. However, because the interpolated dynamics in these plateaus are artificial, we clip the intensities in the final predictions to ±6 for comparable analysis to the ground truth. The window sizes and polynomial orders for the filter were chosen such that the microscale fluctuations are preserved, while the frame-to-frame noise of the system are smoothed.

\subsection{Data analysis}

\textit{No electric field.} In the absence of an electric field ($p=\hat{E}=0$), where $\hat{E}$ is the direction of the homogeneous electric field $\mathbf{E}$, the motion of the boundary is dominated by Brownian motion: the power spectra of the traces closely follow a power law of $1/f^2$ as shown in fig.\ref{fig:data}b,c, without prominent peaks that indicate resonant frequencies. As the dynamics are highly noisy, we choose to represent the traces’ stochasticity via the Hurst exponent $H$, which characterizes the degree of long-term memory within a time series that exhibits a mix of deterministic and stochastic properties \cite{mandelbrot_vanness,hurst,eke}:
\begin{align}
\label{hurst}
\langle(\xi(t+\tau)-\xi(t))^2\rangle_t=\tau^{2H}
\end{align}
Here $\xi(t)$ is some noisy process, and the average is to be taken over many instances of $t$. $H \in (0.5,1]$  indicates positively autocorrelated or persistent events (i.e. long-range memory), and, conversely, $H \in [0,0.5)$ indicates negatively autocorrelated or anti-persistent (i.e. highly volatile fluctuations) events. $H=0.5$ signifies complete absence of long-range memory. $H$ has a direct relationship to a purely noisy process of the fractional Brownian motion type\cite{mandelbrot_vanness}, as we can relate its power law ($P(f)\propto f^{-\gamma})$ to its $H$ by the following:
\begin{align}
\gamma=2H+1
\end{align}
To calculate $H$ from our data, we use the method of rescaled range analysis developed by Hurst for computational efficiency (detailed in ref. \cite{hurst,mandelbrot_wallis}), which has been shown to be equivalent to the relation posed by the structure equation eq. \ref{hurst} \cite{mandelbrot_vanness}. An example computation of obtaining $H$ through this method is shown in fig.\ref{fig:data}d for a representative trace. For our $\hat{E}=0$ data, we find the mean $H$ (averaged over all 400 boundary point traces) is $ H=0.55 \pm .01$, which indeed matches closely with the theoretical value of $H=0.5$ for Brownian motion ($\gamma = 2$) thereby indicating lack of long-term memory. \\

\textit{With an electric field.} With an applied electric field that flips in direction, the external parameter takes the form $p=\hat{E}=\pm 1$, where each sign indicates the relative orientation of $\mathbf{E}$. The cell becomes polarized and collectively moves down the electric potential gradient, while the boundary still exhibits highly stochastic fluctuations. This is shown in fig.\ref{fig:data}a, where approximately half of the cell is now protruding and the other half is retracting, indicating unidirectional collective movement. The fluctuations are still dominantly Brownian, but unlike the case of $\hat{E}=0$, the positive fluctuations \textit{persist} longer in the region of `positive polarization' (protrusions) and likewise for the `negatively polarized' (retracting) region due to the deterministic effects of the external forcing. This should result in increasing our Hurst exponent slightly. In agreement with this, the Hurst exponent of these steady-states under electrotaxis is calculated to be $0.60 \pm .03$. While it is a small increase from that of pure Brownian motion, its value suggests that a reservoir computer should be able to predict the local dynamics for short horizons.\\

\textit{Transient}. Upon flipping the field's direction, the cell's boundary responds gradually as the polarization change locally propagates from one side of the boundary to its diametrically opposed side. This results in the large-scale transient of the polarization that is shown in the kymograph in fig.\ref{fig:data}a. Once the entirety of the cell boundary has changed directions, the cell resumes to collectively migrate down the new potential gradient.

Therefore, we conclude if the reservoir is able to predict (i) the mesoscale transient timescale, (ii) the macroscale steady-state polarizations, and (iii) the power law of the stochastic microscale fluctuations of the boundary, it has correctly learned the dynamics that govern the system.


\section{Results}
\subsection{Global (climate) prediction}

We proceed to measure these three attributes for \textit{fully closed predictions} (i.e. an infinite prediction horizon) as follows. For (i) and (ii), we plot the time series that shows the average location of the positive speeds over the entire boundary for both the prediction and the data. This allows us to visually evaluate whether the prediction was able to predict both the gradual change and the mean `focus' of the polarization. Finally for (iii), we calculate the Hurst exponent of the steady-states of the prediction and compare with that of the data to assess the reservoir's ability to capture the stochastic local fluctuations.

\begin{figure}
    \centering
    \includegraphics[scale=.38]{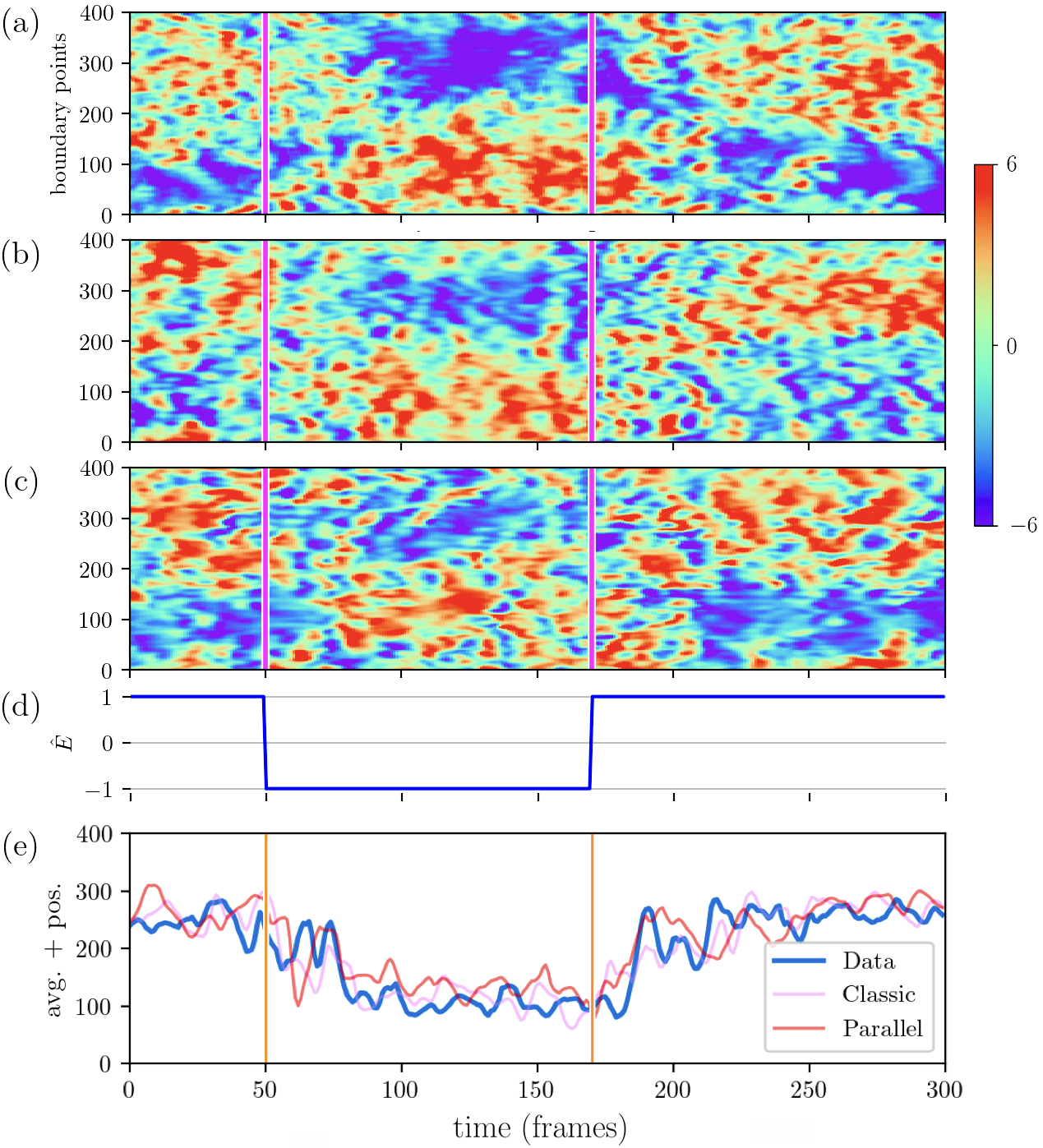} 
    \caption{(a) Boundary motion test data, with switches in $\hat{E}$ demarcated in magenta. (b) Classic (single reservoir) fully closed prediction. (c) Parallel fully closed prediction. (d) Direction of electric field associated with the test data. (e) Average positive positions of data and predictions, weighted by intensities. The $\hat{E}$ switching points are demarcated as orange lines.}
    \label{fig:closed}
\end{figure}

Fig.\ref{fig:closed} displays the fully closed predictions for the classic (i.e. single reservoir) (fig. \ref{fig:closed}a) and parallel architectures (fig. \ref{fig:closed}b), both equipped with the parameter tag (fig. \ref{fig:closed}c). For all predictions, the total number of nodes in the RC (i.e. $\text{N}_\mathbf{r}\cdot \text{N}_R$, where $\text{N}_R$ is the number of reservoirs) has been fixed to 2000. All figures containing the prediction and test data have been shifted by $t-t_0$ (where $t_0$ once again marks the duration of the listening period), such that $t=0$ marks the first prediction, and the listening phase is taken to happen prior to $t=0$. Both classic and parallel reservoirs were able to predict not only the polarization switch of the steady-state, but also the transients that commence as the external parameter tag is switched, despite the sparse number of trained transients. This is evident in fig.\ref{fig:closed}c, which shows the average position of the protruding (positive) region weighted by the speed. The RC's ability to capture the transients as the parameter switches mid-prediction can be seen in the figure. In a dynamical systems framework, the reservoir was able to capture the transition characteristics as it flows from one attractor basin of its phase space (corresponding to the steady state dynamics of one $\hat{E}$ parameter value) to another.

Finally, the Hurst exponents of the steady-states of the classic and parallel predictions were calculated to be $H_\text{classic} = 0.58 \pm 0.04$ and $H_\text{parallel} = 0.60 \pm 0.03$, respectively. Both architectures were able to capture the global climate of the test data ($H_\text{data} = 0.60 \pm 0.03$), demonstrating an RC’s general ability to predict the global statistics of large scale, highly stochastic biological dynamics.

\subsection{Local prediction}
For predictions across a shorter horizon, we turn to a more standard metric of accuracy by using the MAE (mean average error) to assess predictive performance. Because the noise is a random-walk, we may obtain relatively decent predictions for a short horizon before the prediction and data stochastically drift too far from each other, saturating the error. Indeed, our Hurst exponent analysis suggests that the system dynamics, while dominantly noisy, have deterministic elements that should be able to be modeled and predicted at least for short time windows. In turn, the characteristic saturation time of the prediction MAE can be used as a tool for understanding the relative timescales and strengths of stochastic versus deterministic dynamics of the system. We note in passing that by using the MAE, all attributes of the dynamics that were aforementioned in the previous section for the fully-closed predictions are already assessed by nature of the metric.

We normalize the MAE (NMAE) by dividing it by the range of the data, as the time series generally fluctuates around 0. As the prediction error will eventually saturate to its maximum NMAE as the horizon increases, we define a divergence exponent $\lambda$ that characterizes how fast the reservoir's performance saturates by fitting the error with
\begin{align}
\text{NMAE}(t) = M_0(1- e^{-\lambda t})
\end{align}
where $t$ represents the prediction horizon and $M_0$ is the saturated NMAE.

\begin{figure}
    \centering
    \includegraphics[scale=.276]{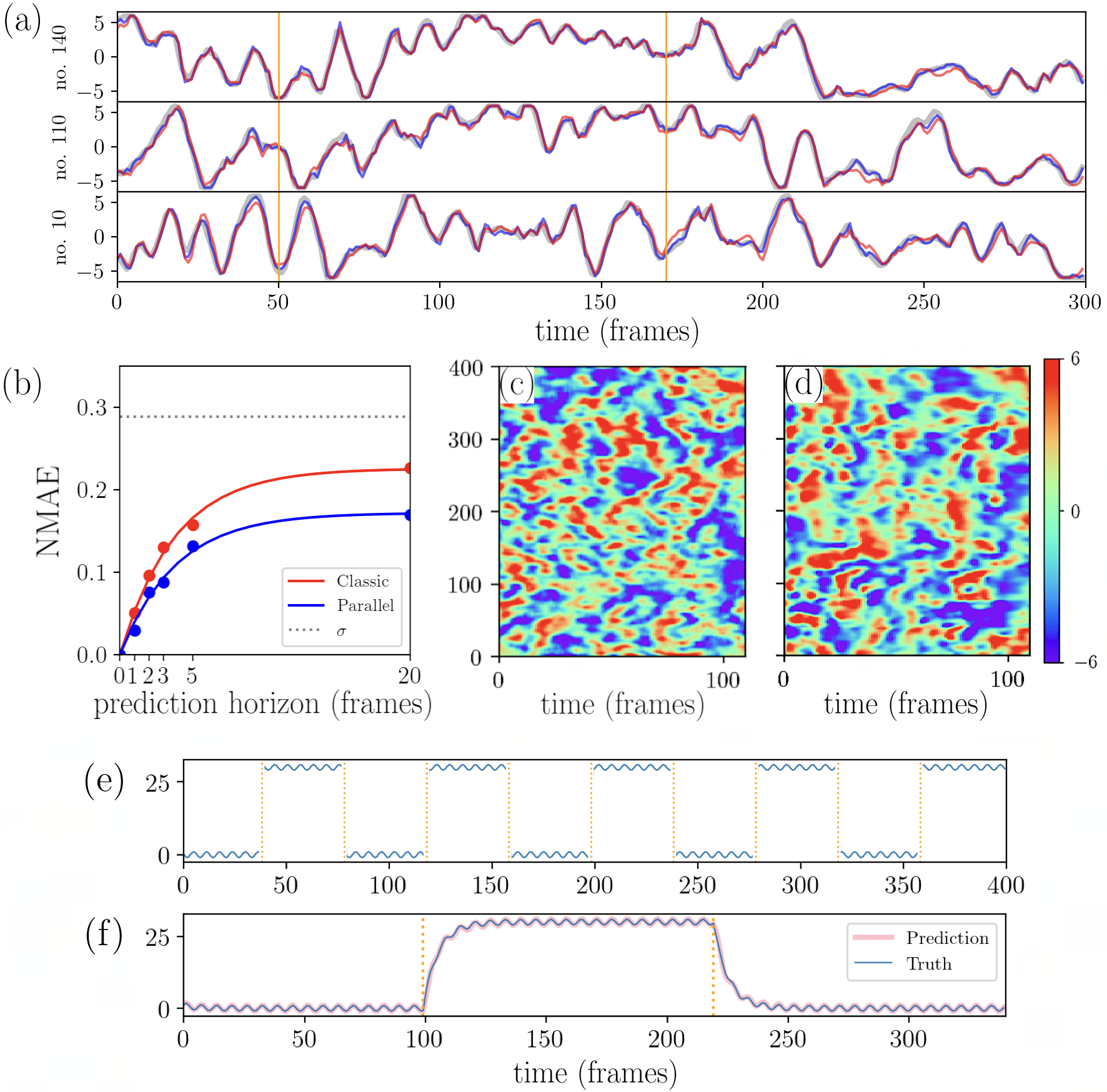} 
    \caption{(a) 1-frame prediction traces of 3 boundary points, with data (gray), parallel prediction (blue), and classical prediction (red),  with $\hat{E}$ switching points demarcated with yellow lines. (b) NMAE as a function of prediction horizons, with standard deviation ($\sigma$) of the entire test data in gray dash ($\approx 3.46$), normalized by dividing it by the data's range. (c) $\hat{E}=0$ data. (d) $\hat{E}=0$ parallel prediction. (e) Steady-state training data of a damped forced circuit, with alternating steady states (demarcated with yellow lines). (f) Prediction of damped forced circuit dynamics with inferred transients}
    \label{fig:local_e0_circuit}
\end{figure}

The divergence time $\tau_\lambda \equiv 1/\lambda$ characterizes the timescale of error saturation. Fig.\ref{fig:local_e0_circuit}a,b displays the results for a few prediction horizons for both the parallel and classic architectures, demonstrating the RC's capability of predicting the local dynamics very well for short prediction horizons in agreement with the values of $H$ obtained previously. The parallel case yields a characteristic time of $\tau_{\text{parallel}}\approx 3.90$ frames with $M_0{_{\text{parallel}}}\approx 0.21$, whereas the classic case gives $\tau_{\text{classic}}\approx 3.35$ frames and $M_0{_{\text{classic}}}\approx 0.23$. While there is slight improvement in predictive power of the parallel architecture for short horizons, the divergence times from both architectures are comparable. This can be interpreted as the relative timescale of the deterministic part of the boundary dynamics, until the stochasticity takes over the system, in agreement with the obtained Hurst value of the data.

We note that these values are still less than the standard deviation of the data, due to the reservoir being able to capture its global attributes.

\subsection{Prediction of new steady states and transients}

\textit{No electric field.} We predict dynamics for $\mathbf{\hat{E}}=0$, which was not trained on, to assess the reservoir’s ability to interpolate unseen dynamics on biological dynamics. The parallel architecture was utilized for this fully closed prediction. We use the same trained $\mathbf{W}_\text{out}$ but set $p=0$. Fig.\ref{fig:local_e0_circuit}c,d demonstrates that the prediction was able to capture the mostly Brownian noise-like dynamics of the system, as evident in the spatiotemporal scales of the boundary fluctuations, without any macroscale polarization. Indeed, $H_\text{prediction} = 0.56 \pm 0.01$, which matches closely to that of the data ($H_\text{data} = 0.55 \pm 0.01$). \\

\textit{Simulated transient.} That RCs can predict transients from sparse data motivates the question of whether the steady-states or parameter values had betrayed any information of the transients, or if the reservoir was able to interpolate the transients from the sparse transient observations alone. While studies have been conducted on the ability of RCs to anticipate transients in non-steady chaotic systems by providing training data of parameter values neighboring the transient regime, the mean lifetime of these transients obeyed a close relationship with the external parameter \cite{transient_powerlaw,chaotic_transient}. Therefore, with ample training data of steady states corresponding to neighboring parameter values, the reservoir had information to assist in the interpolation or extrapolation to this unseen regime.

We proceed to investigate the extent to which a reservoir can learn transients from sparse data by considering a simple linear system, which makes it easy to equip it with transients independent of steady-states. We consider a single time series $Q(t)$ that evolves as a charging or discharging capacitor in a damped circuit, equipped with a simple periodic forcing:
\begin{align}
\label{charge1}
  \renewcommand{\arraystretch}{1.2}
  \cos(t)-\tau \sin(t) = \left\{
    \begin{array}{@{\,}l@{\,\,\,}l@{}}
      Q_0-\tau\dot{Q}-Q & \text{\small(charge)}\\
      -\tau\dot{Q}-Q  & \text{\small(discharge)}
    \end{array}
    \right.
\end{align}

\noindent where $Q_0$ is some constant that supposes the maximum charge in the absence of forcing, and $\tau$ is the characteristic time of the transient. 

The solutions to these equations (with $Q_{\text{charging}}(0)=0$ and $Q_{\text{discharging}}(0)=Q_0$) are given by:
\begin{align}
\label{charge2}
  \renewcommand{\arraystretch}{1.2}
  Q(t) = \left\{
    \begin{array}{@{\,}l@{\,\,}l@{}}
      e^{-t/\tau}(1-Q_0)+(Q_0-\cos(t)) & \text{\footnotesize(charge)}\\
      e^{-t/\tau}(1+Q_0)-\cos(t)  & \text{\footnotesize(discharge)}
    \end{array}
    \right.
\end{align}

\noindent Any information about the timescale of the transients, $\tau$, is removed from the steady-state; a reservoir trained purely on the steady-states should have no knowledge of them.

We generate a training dataset with just the steady-states, and tag the discharged and charged states with $p=+1$ and $-1$, respectively (fig.\ref{fig:local_e0_circuit}e). $Q_0$ and $\tau$ were arbitrarily chosen to be 30 and 6, respectively.  We let the reservoir predict the test data, which contains the transient. As seen in fig.\ref{fig:local_e0_circuit}f, the reservoir can interpolate the timescale of the transient. Because this is a fully closed prediction, the reservoir has seen the transient only when it was synchronizing with the test data during the listening stage. This suggests that the reservoir has been able to learn not only the steady-states corresponding to the external parameter value, but also their transients, from sparse observations either in the training data \textit{or} the test data, with a good choice of hyperparameters. In a dynamical systems framework, the reservoir is able to infer how to transition between coexisting basins of attractions in its phase space from very sparse observations.\\


\section{Conclusion}
We demonstrated that RCs can learn complex dynamics of different parts of the phase space with very short time series. We tested its efficacy on highly stochastic biological data with very little information on the underlying dynamics, and have demonstrated that the reservoir responds dynamically to a switch of the parameter value. We have shown that reservoir computers can decently predict the true time series of the system across short horizons, with improvement in prediction accuracy via a parallel scheme. The saturation time of the error for local predictions provides important information on the relative timescales of the deterministic and stochastic elements of the system dynamics. RCs can interpolate various types of dynamics as well, such as transients and unseen parameter values, from very sparse observations, allowing them to act as dynamic twins of complex dynamical systems.

\begin{acknowledgments}

We would like to thank Michelle Girvan for helpful discussions. This material is based upon work supported in part by the Air Force Office of Scientific Research under award number FA9550-21-1-0352 and Army Research Laboratory grant W911NF2320040.  

\end{acknowledgments}

\nocite{*}

\bibliography{main}

\end{document}